\documentclass[conference]{IEEEtran}
\IEEEoverridecommandlockouts
\usepackage{lmodern}
\usepackage{cite}
\usepackage{amsmath,amssymb,amsfonts}
\usepackage{algorithmic}
\usepackage{graphicx}
\usepackage{textcomp}
\usepackage{xcolor}

\def\BibTeX{{\rm B\kern-.05em{\sc i\kern-.025em b}\kern-.08em
    T\kern-.1667em\lower.7ex\hbox{E}\kern-.125emX}}
\begin{document}

\title{Extended Event Log: Towards a Unified Standard for Process Mining\\
}

\author{\IEEEauthorblockN{Ali Suleiman}
\IEEEauthorblockA{\textit{Department of Business Informatics} \\
\textit{German University in Cairo}\\
Cairo, Egypt \\
ali.midhat@guc.edu.eg}
\and
\IEEEauthorblockN{Gamal Kassem} \IEEEauthorblockA{\textit{Department of Business
Informatics} \\
\textit{German University in Cairo}\\
Cairo, Egypt \\
gamal.kassem@guc.edu.eg} }

\maketitle

\begin{abstract}
    Process mining has grown popular today given their ability to provide
    managers with insights into the actual business process as executed by
    employees. Process mining depends on event logs found in process aware
    information systems to model business processes. This has raised the need to
    develop event log standards given that event logs are the entry point to any
    process mining project. One of the main challenges of event logs and process
    mining in general as was mentioned by the IEEE task force on process mining
    deals with the finding, merging \& cleaning event data. This resulted in
    having multiple event log standards with different features. This paper
    attempts to propose a new unified standard for event logs that enriches the
    results of process mining without the need to tailor event logs for each
    process mining project.
\end{abstract}


\section{Introduction}

All businesses rely on a sequence of tasks to add value, collectively known as
the business process. Within the intricate fabric of a business, these processes
play a vital role in its functioning and profit generation \cite{1}. The
business process stands as a cornerstone, ensuring the success of an
organization when managed effectively by its leaders.

Building upon this understanding of the pivotal role of business processes, a
new discipline emerged to reveal the complexities within these sequences of
tasks. This discipline, named process mining, delves into the analysis of event
data to monitor, control and enhance business processes. Process mining provides
a lens through which organizations can gain valuable insights into the actual
execution of processes, allowing managers to take more informed decisions, that
actually enhances and improves the organization \cite{2}. For this purpose,
process mining is important as it reveals the actual process carried out by
employees, which may deviate and differ from the actual process \cite{1}.

Process mining rely on data available from event logs, and while there are many
available event log standards the challenge first mentioned by the IEEE task
force on process mining remains unanswered which is how to collect merge and
clean event data \cite{3}. Thus, this paper aims to answer the question of how
to develop new unified event log standards that do not require tailoring an
event log for the process mining project but is rather generic and could be used
for any project.

\section{Preliminaries}
In this section, the basic concepts and terminologies used in this paper will be
defined and explained. These concepts are essential to understand the rest of
the paper.

\subsection{Process Mining}
With the introduction of Process Aware Information Systems (PAISs), the concept
of process mining was enabled. Process mining aims to extract the execution of
business process as recorded in the event logs of PAIS, to construct as- is
business process models \cite{4}. This automatic construction is needed because
available models are often idealized and do not reflect the actual process
carried out by employees \cite{5,2}.

Process mining aims at discovering, monitoring and enhancing business processes.
This is achieved through the three different process mining types:

\begin{itemize}
    \item Process Discovery
    \item Process Conformance Checking
    \item Process Enhancement
\end{itemize}

The first and most used type of process mining is process discovery. The goal in
process discovery is to discover the as- is process models which are being
executed by the employees based on the event log data found in PAISs \cite{6}.
In process discovery, the process model should be discovered automatically
without prior knowledge of the business process \cite{6,7}. Thus, the input to
process discovery is an event log and the output is a process model \cite{3}.

The second type of process mining is conformance checking. The goal in
conformance checking is to compare an existing process model (from the business
documents) to the event log to discover areas of deviations \cite{8}.
Accordingly, conformance checking takes as input both a process model and an
event log, to output diagnostics \cite{6,3}.

Finally, the third type of process mining is process enhancement. The goal of
process enhancement is to output a new, enhanced, process model that is better
aligned and matches the as-is process executed by the employees. Thus, the input
to process enhancement is an existing process model and an event log while the
output is a new model that is more aligned with reality \cite{9}. Process
enhancement could also enhance the process itself; it doesn't have to only
change the model. In such cases, process enhancement is used to identify areas
that could be enhanced such as bottlenecks. This means that the output model
could be an extended model that is more optimized, increasing the efficiency of
the business process \cite{5,6,9,10}.

\subsection{Event Logs}

As discussed in the previous section, event logs are the entry point to the
three types of process mining and thus are essential to process mining projects.
In fact, process mining is made possible due to the availability of event logs
found in PAISs \cite{6,3}. This section will focus on event logs, it will start
by defining what event logs are followed by a discussion of the essential
requirements that should be made available in event logs.

Event logs are log files made available by PAIS that record the execution of
activities that are the result of the interaction with the system. It also
records data such as case id, timestamp, resources involved alongside other
relevant information about the recorded task. This data allows the analyst to
gain insights into the actual executed process in terms of performance and
compliance \cite{2}.

While the term event log implies that it is readily available for process
mining, this is not the case. Although \cite{2} stated that event logs should be
treated as first class citizens, PAIS usually just generates a series of logs
that could be utilized to build event logs. However, event logs could also
capture data from other sources such as mail archives or message logs \cite{4}.

The first challenge of process mining as was mentioned in the process mining
manifesto, is “finding, merging \& cleaning event data”. This challenge is
concerned with extracting suitable process mining data, which can be difficult
and time-consuming. This is because the data should be relevant to the context
of the process mining project, of high quality, of the same granular level,
among other criteria. The data could also be available in different sources.
Also, the quality of the process mining project depends on the quality of the
extracted data \cite{2}.

\section{Literature Review}

In this section we conduct a literature review on event logs
to identify the existing standards and compare their structure
in order to identify the research gap in the area of unifying the
event log standards for process mining. The first part presents
the state of the art while the second part discusses the findings
of the state of the art so that the research gap could be
scientifically extracted.

\subsection{State of the Art}
\subsubsection{eXtensible Event Stream (XES)}

Extensible Event Stream (XES) was first introduced in
2010 by the IEEE taskforce on process mining. XES saw great
success and was very important for process mining,
considering that it was one of the first efforts toward event log
standardization \cite{11}.

A process log and its trace are defined by their elements,
which are events. Events are classified using a set of attributes.
Extensions are defined in terms of concept, life cycle,
organization, time, and semantics. They are used to define a
set of attributes on any level of the XES log hierarchy \cite{12}.

The XES is an XML based format for collecting event
logs. It has been adopted as the default format for
interchanging event log data by the IEEE taskforce on process
mining \cite{3}.

Although XES saw a great success as process mining grew
popular and event logs got more complex, they soon presented
a challenge, namely the convergence and divergence problems
\cite{13}. These problems rose because of the single case notation
of XES, where each event is related to a case \cite{14}. The
convergence problem is defined by \cite{9} as the repeated
occurrences of the same group of activities within a single
case. While convergence is defined as the relating of one event
to multiple cases.

To overcome the challenge of the single case notation,
present in the XES format a new format was proposed, namely
the Extensible Object Centric Event Log (XOC), which will
be discussed in the following section.

\subsubsection{Extensible Object Centric Event Log (XOC)}

XOCs are the first attempt in including multiple case
notations in process mining data. The concept of XOC was
based on the idea that an event log has a list of events, each of
which refers to an object model \cite{15}. Where an object
represents the appearance of an active player in the process
such as customer, invoice, order, etc. \cite{19}.

XOC relates events to objects instead of cases, as is the
case XES, which solves the single case notation problem
which enriches the presentation of event data. Additionally, it
can better deal with one-to-many and many-to-many
relationships. These contributions enrich the results of process
mining projects by revealing the complex interactions
between the behavioral and data perspectives \cite{15}.

On the other hand, this format suffered from bad
performance because the log scaled badly with the number of
events and objects involved in the process. This is mainly
since in XOC the objects are replicated for all the events that
are correlated with the object \cite{11}.

To overcome the performance issues of the XOC a new
format was proposed, Object Centric Event Log (OCEL),
which will be discussed in the following section

\subsubsection{Object Centric Event Log (OCEL)}

OCELs are similar to XOCs where they also drop the
single case notation of the XES format and adopt the object
notation. On the other hand, they propose different structures
to overcome the performance issues of XOC \cite{16}.

OCEL contains data about the process such as events,
objects, timestamps and other data. It also contains global log,
global event and global object, which describe the log, events
and objects respectively. The reasoning behind these global
fields is to describe what should be available in the log, events
and objects included in the OCEL. This is the feature that
ensures that objects or events won't have to be replicated to
overcome the XOC performance issues \cite{11}.

An OCEL contains global log, global event and global
object. These contain data about the log, event and object
respectively. event can have a list of objects referencing the
objects involved in each activity. This then references the
actual object where the information about the object is actually
stored. This schema ensures that the information about the
objects won't be repeated for each event, reducing redundancy
and optimizing the log \cite{11,14}.

The challenge of OCEL, however, was that it had to be
flattened into an XES that support the single case notation in
order to be visualized using the traditional process mining
tools [17]. There have been multiple efforts directed towards
flattening the OCEL dynamically based on the analysis to
generate a log that can be visualized based on the chosen
business object \cite{18}.

OCEL, like other event logs, has to be manually collected
and tailored for the specific process mining project needs \cite{3}.
Another approach for standardizing event log repositories for
process mining is process warehouse, which will be discussed
in the next section.

\subsubsection{Process Warehouse}

The concept of process warehouses enables what is known
as Multidimensional Event Log (MEL) or Process Cubes.
Process warehouses are similar to traditional data warehouses,
but instead they contain process data. It could be then used
with Online Analytical Processing (OLAP) operators to
conduct process analysis \cite{20}.

The purpose of the warehouse is to consolidate data across
different sources into a single centralized trusted repository.
The warehouses are used for their low load-times, fast
response times and ability to handle ad-hoc queries. \cite{19,21}.
It includes a fact table that usually contains measurable
numerical values used to analyze the performance of the
business process, while the dimensions contain description of
the business process \cite{21}.

Similar to OCEL, process warehouses can contain
business objects and relate them to the activities. In a proposal
by \cite{19}, they also include a case notation as shown in figure 1
below. Multiple proposals such as \cite{10} and \cite{11}, utilize
process warehouses to extract and materialize OCELs.

The challenge of process warehouses, however, is that
they do not support the direct visualization of process models,
instead visualization and process mining is done on a
materialized log that is the result of the OLAP analysis.
Accordingly, it hinders the ability to have an interactive
analysis of the process data \cite{20}. They also require an extract,
transform and load (ETL) process, which can prove to be as
complex and time-consuming as the construction of an event
log \cite{21}.

This section presented the state of the art of different event
log standards used in the recent literature; the following
section will contain a discussion of the results presented.

\begin{figure}[htbp]
    \centerline{\includegraphics[width=2.5in]{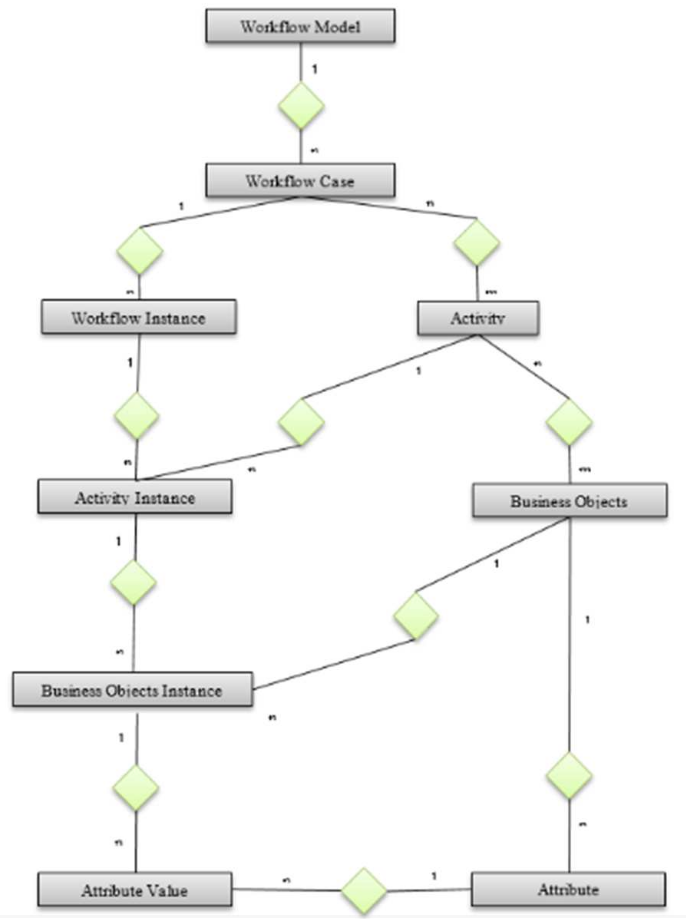}}
    \caption{Workflow Model \cite{19}.}
    \label{fig}
\end{figure}

\subsection{Discussion}

In light of what was discussed in the previous section. We can see that there is
a lot of development in the collection and storage of process data and event
logs development domain. While we see that there are multiple formats and
standards, each having its advantages and challenges, we can conclude that there
are no unified standards for the storage of event data.

\begin{table}[htbp]
    \caption{Feature Comparison}
    \begin{center}
        \begin{tabular}{|l|c|c|c|c|}
            \hline & XES & XOC & OCEL & \begin{tabular}{c} 
            Process \\
            Warehouse
            \end{tabular} \\
            \hline Cases & $\checkmark$ & X & X & $\checkmark$ \\
            \hline Business Objects & X & $\checkmark$ & $\checkmark$ & $\checkmark$ \\
            \hline \begin{tabular}{l} 
            Tailored for Each \\
            Process Mining \\
            Project
            \end{tabular} & $\checkmark$ & $\checkmark$ & $\checkmark$ & X \\
            \hline \begin{tabular}{l} 
            Have Meta Level \& \\
            Instance Level
            \end{tabular} & X & X & $\checkmark$ & $\checkmark$ \\
            \hline \begin{tabular}{l} 
            Can Analyze Instance \\
            Level
            \end{tabular} & X & $\checkmark$ & $\checkmark$ & $\checkmark$ \\
            \hline \begin{tabular}{l} 
            Has Different Levels \\
            of Granularity
            \end{tabular} & X & X & X & $\checkmark$ \\
            \hline Directly visualized & $\checkmark$ & X & X & X \\
            \hline
        \end{tabular}
    \end{center}
\end{table}

Shown in table 1, are the feature comparison of the
different event log standards discussed in the state-of-the-art
section. The cases feature is concerned with whether the
standard supports having cases and case id. Business Objects
is concerned with the support of relating events to business
objects. Tailored for each process mining project is concerned
with the ability of the standard to be generic and
comprehensive to any process mining project, meaning that it
doesn’t have to be built and tailored for different process
mining projects. The meta and instance level are concerned
with the structure of the event log standard, where, if it has
meta and instance level, this means that the event log could be
meta described to allow it to be generic and eliminate
duplications. Having the ability to analyze instance levels
means that different specific instances could be selected for
analysis showing the different business objects involved.
Having different levels of granularity means that the log can
describe the task and the task steps (what happened for a task
to be completed) could be described. Finally, directly
visualized means that the log could be directly visualized by
traditional process mining tools.

The results of the comparison clearly show that process
warehouses have most of the features needed in a unified
event log standard. However, unlike XES it couldn't be
directly visualized, but rather a sub log must be materialized
based on the OLAP analysis \cite{20}. It also has the challenge of
requiring an ETL process and thus can be complex to build.

Thus, there is a clear gap where a new event log should be
developed that enables the generic comprehensive
representation of event data, so that an event log doesn't have
to be tailored and built for every process mining project as
well as eliminates the need for a complex ETL process.

\section{Methodology}

To address the previously stated gap, which was identified
in the previous section, a methodology based on the design
science research methodology is utilized. The methodology is
chosen due to its iterative nature and its focus on creating an
artifact that solves the problem \cite{22}.

The methodology constitutes the following 6 steps:
\begin{enumerate}
    \item Problem Identification: The problem is identified based on the
    conducted literature review.
    \item Objectives: The objectives of the solution are identified and
    logically inferred from the previous step.
    \item Design and Development: The artifact addressing the problem and
    objectives identified is developed.
    \item Demonstration: The artifact is used to demonstrate how it solves the
    problem.
    \item Evaluation: The artifact is evaluated based on how well it sponsors
    the solution to the problem.
    \item Communication: The designed artifact and the problem it solves, is
    communicated to the relevant stakeholders.
\end{enumerate}

\section{Results}

This section will discuss the results of conducting the
previously mentioned methodology in order to develop new
event log standards that address the research gap of this paper.

\subsection{Problem Identification}

In the first phase of the methodology the problem has to
be identified. We identify the problem based on the literature
review conducted in the previous section. The problem of this
paper is the lack of a readily available generic and
comprehensive event log that can analyze the task and task
step of a process while having the ability to be directly
visualized. Where the generic property of the log implies that
the log can describe any process regardless of the domain and
the comprehensive property means that it contains all data and
doesn't have to be tailored for each process mining project.

\subsection{Objectives}

The objectives of this paper, based on the problem
identification, are to develop a new event log format that is 1)
generic, 2) comprehensive and 3) includes the different
granular levels.

\subsection{Design and Development}

\begin{figure}[htbp]
    \centerline{\includegraphics[width=2.5in]{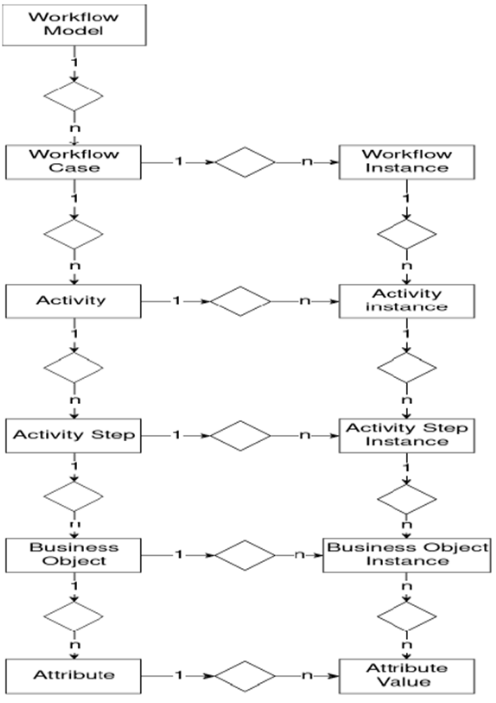}}
    \caption{Proposed Extended Event Log Adapted from \cite{19}.}
    \label{fig2}
\end{figure}

\begin{figure*}[!t]
    \begin{center}
        \includegraphics[width=5in]{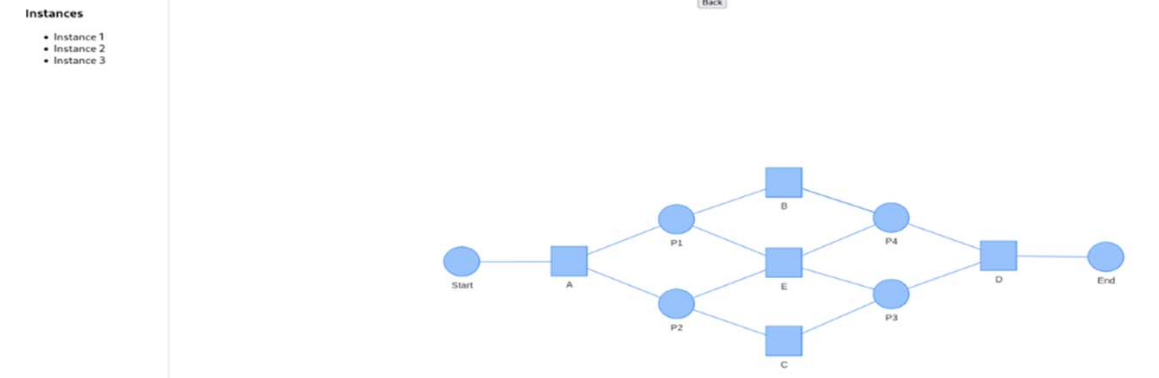}
        \caption{Petri Net Output from the Proposed Event Log.}
    \end{center}
\end{figure*}

To design the newly proposed event log, we consider
previous event log formats and structures. We find that event
logs have common elements as was discussed by \cite{23}, those
are the id, activity and timestamp. Thus, those elements should
be present in our event log.

Since the goal is to have an event log structure similar to
what is present in process warehouses, we chose the structure
proposed by \cite{19} to be extended to fulfill our requirements.
Studying figure 1, we see that the model includes an instance
level and a meta level where the log is meta described. The
meta description could be used for the visualization of the
process model while the instance level could be used to
analyze specific instances of the process and how they relate
to the different business objects involved. However, the model
is missing the task steps involved, which should be taken into
consideration in our model.

Shown in figure 2, is our proposed event log format. Our
model adds to the model proposed by \cite{19}, where it adds an
activity step with a one-to-many relationship with activity.
This means that each activity can have multiple activity steps.
Business objects have also been moved to relate to activity
steps instead of activity. This means that each activity step has
multiple business objects involved, those can be the user, the
screens or other active players. The activity step also has an
instance and a meta level to prevent having duplicates.

\subsection{Demonstration}

For demonstration purposes, an example event log has
been developed in XML format. A tool that reads such an
XML file was also developed based on the alpha miner
algorithm in order to output a petri net. In the developed tool,
clicking on a transition, the tool shows the task steps involved
to mark the task as completed. It also shows a list of instances,
where clicking on an instance highlights the route of the
specific selected task.

This tool is still in a primary phase and has been only used
as a proof of concept to demonstrate the event log.

Given that this research is still a work in progress and this
paper is intended to be primary results, no evaluation has been
conducted as of now, and we acknowledge that this is a
limitation for this work.

\section{Conclusion}

In conclusion, this paper proposed a new format for an
extended event log based on XML that is comprehensive,
generic and readily available for analysis if adopted by IS
systems. This new format overcomes challenges presented in
traditional event logs, where it won't necessarily need to be
tailored for each process mining project, while showing more
data about the event data such as the business objects
involved, and the task steps involved.

However, we do acknowledge some limitations of this work, namely that it is
still work in progress and thus hasn't been properly evaluated by experts in the
field of process mining. In future work, it should also be demonstrated on a
real-life case to see how it performs under real conditions. The tool developed
is based on the alpha miner algorithm which is known to perform poorly in real
environments and thus, different algorithms should be used and tested with this
event log format. The tool, however, is part of a bigger project and is not the
focus of this paper but was used for demonstration purposes.

Overall, this work proposes a new event log standard that
could potentially enrich the results of process mining. It is also
considered to be a step towards addressing the first process
mining challenge as was mentioned by \cite{3} in the process
mining manifesto which deals with the finding, merging \&
cleaning event data.

\bibliographystyle{IEEEtran}
\bibliography{bibl}

\end{document}